\documentstyle[11pt,newpasp,twoside,epsf]{article}

\def\edcomment#1{\iffalse\marginpar{\raggedright\sl#1\/}\else\relax\fi}
\reversemarginpar

\begin{document}

\title{The temporal characteristics of the Chandra X-ray Observatory high energy particle background}
\author{Catherine E. Grant}
\affil{Center for Space Research, Massachusetts Institute of Technology, Cambridge, MA 02139}

\author{Mark W. Bautz}
\affil{Center for Space Research, Massachusetts Institute of Technology, Cambridge, MA 02139}

\author{Shanil N. Virani}
\affil{Harvard-Smithsonian Center for Astrophysics, Cambridge, MA 02138}

\begin{abstract}
It was observed early on in the Chandra X-ray Observatory mission 
that the background rates of the Advanced CCD Imaging Spectrometer 
(ACIS) were highly variable on both short and long timescales.  We 
present analysis of lightcurves of the high energy 
($>$ 15 keV) ACIS background spanning most of the mission lifetime.  
These high energy events are not produced by astrophysical X-rays, but by the 
particle background, primarily energetic protons.  
Temporal characteristics of both the quiescent background and 
background flares will be discussed as well as correlations with 
Chandra's orbital position and the solar activity cycle.  The strength 
and frequency of background flares appear cyclic with a quasi-annual 
period, most likely from the motion of Chandra's orbit through the 
geomagnetic environment.
\end{abstract}

The Chandra X-ray Observatory, the third of NASA's great observatories in space, was launched just past midnight on July 23, 1999, aboard the space shuttle {\it Columbia}.  After a series of orbital maneuvers Chandra reached its final, highly elliptical, orbit.  Chandra's orbit, with a perigee of 10,000~km, an apogee of 140,000~km and an initial inclination of 28.5\deg, transits a wide range of particle environments, from the radiation belts at closest approach through the magnetosphere and magnetopause and past the bow shock into the solar wind.

The Advanced CCD Imaging Spectrometer (ACIS), one of two focal plane science instruments on Chandra, utilizes charge-coupled devices (CCDs) of two types, front- and back-illuminated (FI and BI).  The BI CCDs are more sensitive at low X-ray energies, but are also less efficient at rejecting background events and more sensitive to background flares.  Soon after launch it was observed that the background rates on ACIS were highly variable on both short and long timescales.  Understanding the temporal characteristics of the ACIS particle background will allow better modeling and removal of the instrumental background from astrophysical data.

\section{Measuring ACIS background}

\begin{figure}
\plotone{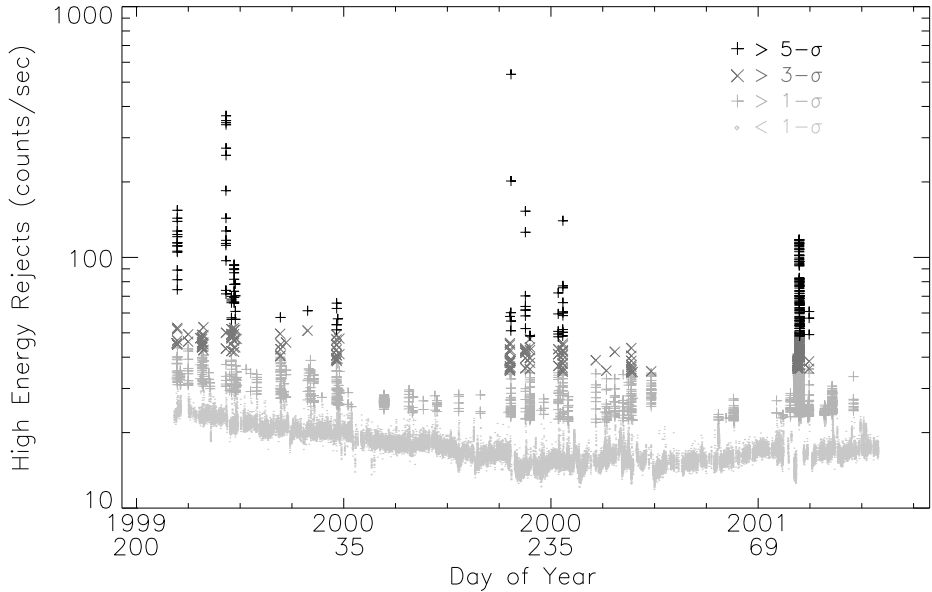}
\caption{Lightcurve of the S3 high energy reject rate from 23 August 1999 through 03 July 2001 in 5 minute bins.  The greyscale and symbols indicate when the background is more than 1-, 3- and 5-sigma above a polynomial fit to the quiescent level.}
\end{figure}

While the ACIS background can be studied with standard X-ray event lists, analysis is complicated by interference from astrophysical X-ray sources which may be extended or variable and must be carefully removed.  We are instead using count rates of events with large pulseheights ($>$ 3750 ADU or $\sim$ 15~keV).  These events are rejected by the flight software and not telemetered to the ground to reduce telemetry bandwidth.  Only the number per frame is included in the exposure records without any information about position or grade (the `DROP\_AMP' field in the level 1 file *\_stat1.fits).  At these high energies, the effective area of Chandra's mirrors is essentially zero, so the events are caused by energetic particles and not astrophysical X-rays.  S3 high energy rates are preferred because the BI CCD is relatively unaffected by changing CTI and radiation damage and is more sensitive to changes in the background flux.  The S3 high energy reject rates are correlated with both the telemetered event rates and the EPHIN radiation monitor on board Chandra (Plucinsky \& Virani 2000).

Figure~1 is a light curve of the S3 high energy reject rate in 5 minute bins from 23 August 1999 when the high energy filter was first activated until 03 July 2001.  The data include all ACIS observations taken in standard faint mode (no window filter or subarray), with no regard to the instrument configuration (i.e. SIM or spectroscopic grating position).  The total exposure time is $\sim$20 Msec.  The background rate is clearly variable on both short and long time scales.  The greyscale and symbols indicate when the background is more than 1-, 3- and 5-sigma above a polynomial fit to the quiescent level.

\section{Quiescent Background}

\begin{figure}
\plotone{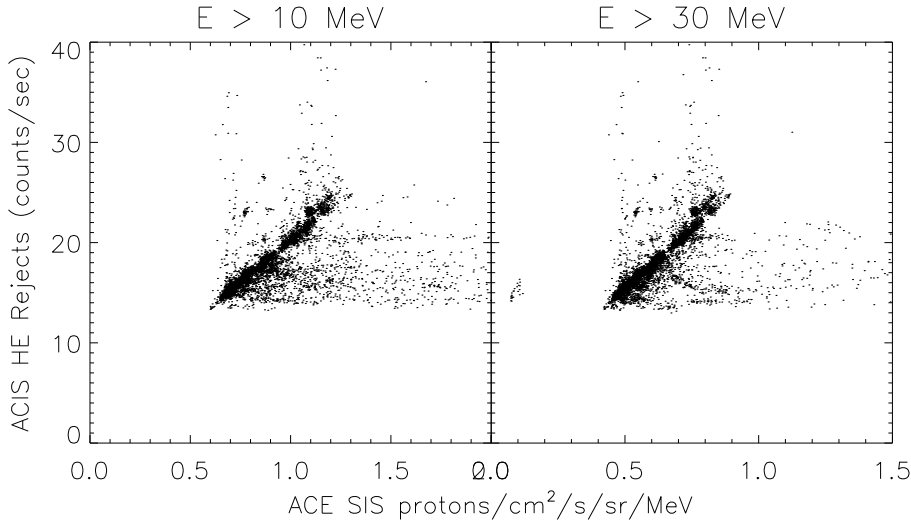}
\caption{Comparison of ACIS background to the proton flux from ACE SIS in the E~$>$~10~MeV and E~$>$~30~MeV bands.}
\end{figure}

As seen in the high energy background lightcurve, the count rate of the quiescent background declined from launch until late 2000 and appears to now be slowly recovering.  This is also seen in the telemetered event background rate (see Chandra X-ray Observatory Center Calibration web pages).  To better understand this temporal structure we compare the ACIS high energy background to the flux of protons measured by the Solar Isotope Spectrometer (SIS) on the Advanced Composition Explorer (ACE) spacecraft (Stone et al. 1998).

Figure 2 compares the ACIS background to the proton flux measured by ACE SIS with E~$>$~10~MeV and E~$>$~30~MeV in hourly bins.  Overall the rates are well correlated.  The streaks going up and to the right are from background flares seen by one instrument only.  That the flares are not concurrent is expected since ACE is located at the Sun-Earth L1 point while ACIS is in Earth orbit.

During solar quiet times, SIS rates are dominated by cosmic rays.  The quiescent high energy background is therefore a reasonable measure of the cosmic ray rate incident on the CCDs. The cosmic ray background measured by ACE and ACIS has declined and then recovered since Chandra's launch.  This change is due to the increase and decrease of magnetic field strength in the heliosphere associated with the 11-year solar cycle.  Now that solar maximum is passing, the ACIS cosmic ray background should continue to level out and increase. 

\section{Flaring Background}
The light curve of the ACIS background includes many ``flares'' or short term increases in
the count rate (Plucinsky \& Virani 2000).  These flares can increase the high energy
reject rate by as much as a factor of 50 and last anywhere from minutes to hours.  The same flares are also seen in the telemetered event rate and can even cause telemetry saturation.   While the origin of the flares is not completely understood, XMM has observed similar events and has determined that they are caused by low energy ($\la$ 100 keV) protons (Str\"{u}der et al. 2001).  

We would like to have a better understanding of what causes these background flares for the purposes of better mission planning, to help us model the Chandra radiation environment, and to better model and remove the instrumental background from astrophysical data.  A number of factors may determine whether a flare occurs including Chandra's position relative to the Earth's radiation belts and magnetosphere, and the level of solar activity.

\subsection{Frequency of Flares versus Instrument Configuration}

\begin{figure}
\plotone{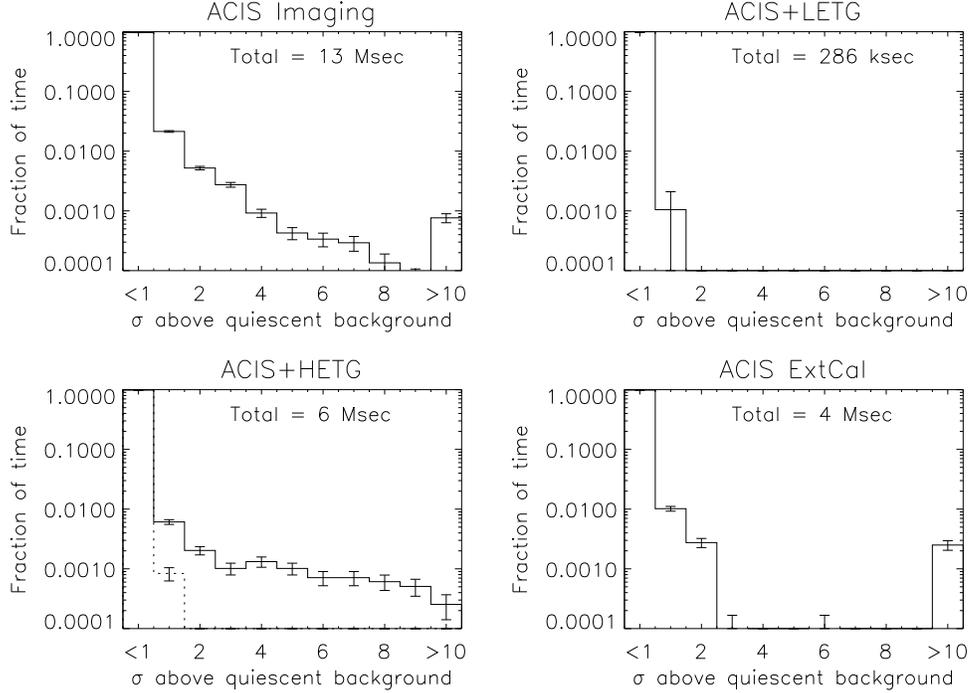}
\caption{Fraction of time spent in each instrumental configuration in which the  ACIS background is above the quiescent background.}
\end{figure}

While the high energy reject rate alone is insufficient to determine the energy of the background particles, examining the relative frequency and strength of flares in different instrumental configurations may provide some clue.  Figure~3 shows the fraction of time spent in each instrumental configuration in which the background was higher than the quiescent level.  Each configuration represents a increasing level of absorbing material between the detector and the sky, from the ACIS imaging configuration, then adding the transmission gratings (LETG and HETG, respectively), and finally to the calibration position when ACIS is out of the focal plane entirely.

For all detector configurations the amount of time impacted by substantial flares is small, of order a few percent.  The flaring probability for ACIS+HETG is completely dominated by a single event on 19 April 2001 associated with an X14-class solar flare and an Earth-directed coronal mass ejection.  The dotted line shows the ACIS+HETG flaring probability without the affected time period.  After this correction, flares are clearly more common in the ACIS imaging position than in the other configurations implying that most background flares are from weakly penetrating particles.  The remainder must be from harder events that penetrate even the detector shielding; most of these can be associated with specific solar events. 

\subsection{Frequency of Flares versus Altitude}

\begin{figure}
\plotone{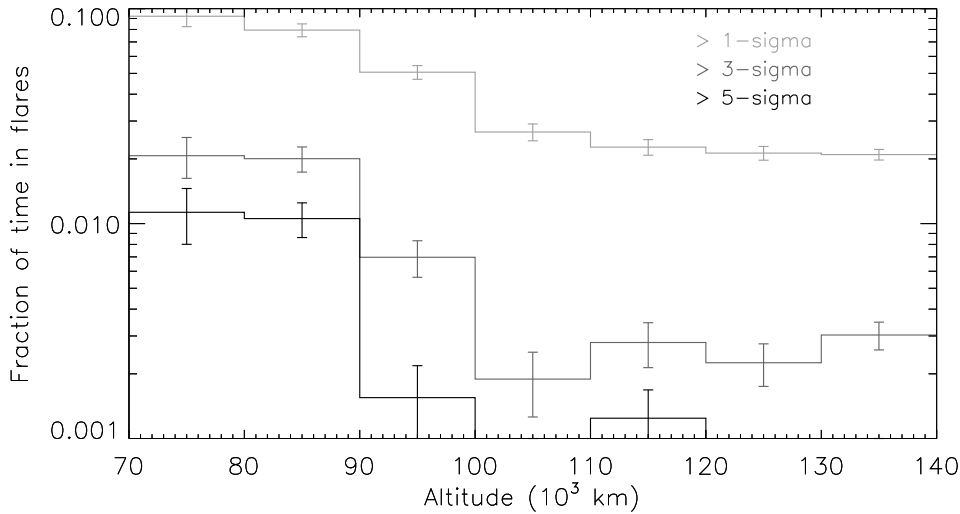}
\caption{Fraction of time at a given altitude that the ACIS background is above the quiescent level.}
\end{figure}

Chandra is in a highly elliptical orbit and once every 2.6 days passes through the Earth's radiation belts.  During this time, ACIS is in a protected position out of the telescope focus.  Observations resume when the radiation model used by the mission planners indicates the particle flux is low enough, usually around an altitude of 60,000 km, however during periods of high geomagnetic activity the size of the radiation belts can increase substantially.  If background flares are associated with the edge of the radiation belts, their frequency and strength should be highly correlated with altitude.  Figure~4 shows the fraction of time at any given altitude that the background is above the quiescent level, using the same color scheme as Figure~1.  Only ACIS imaging observations are included to remove dependence on the particle energy spectrum.  There is some correlation between Chandra's altitude and flare frequency; flares are three times as likely below 100,000 km as above, however even at apogee there is an significant probability of background flaring.

\subsection{Frequency of Flares versus Orbital Position}

\begin{figure}
\plotone{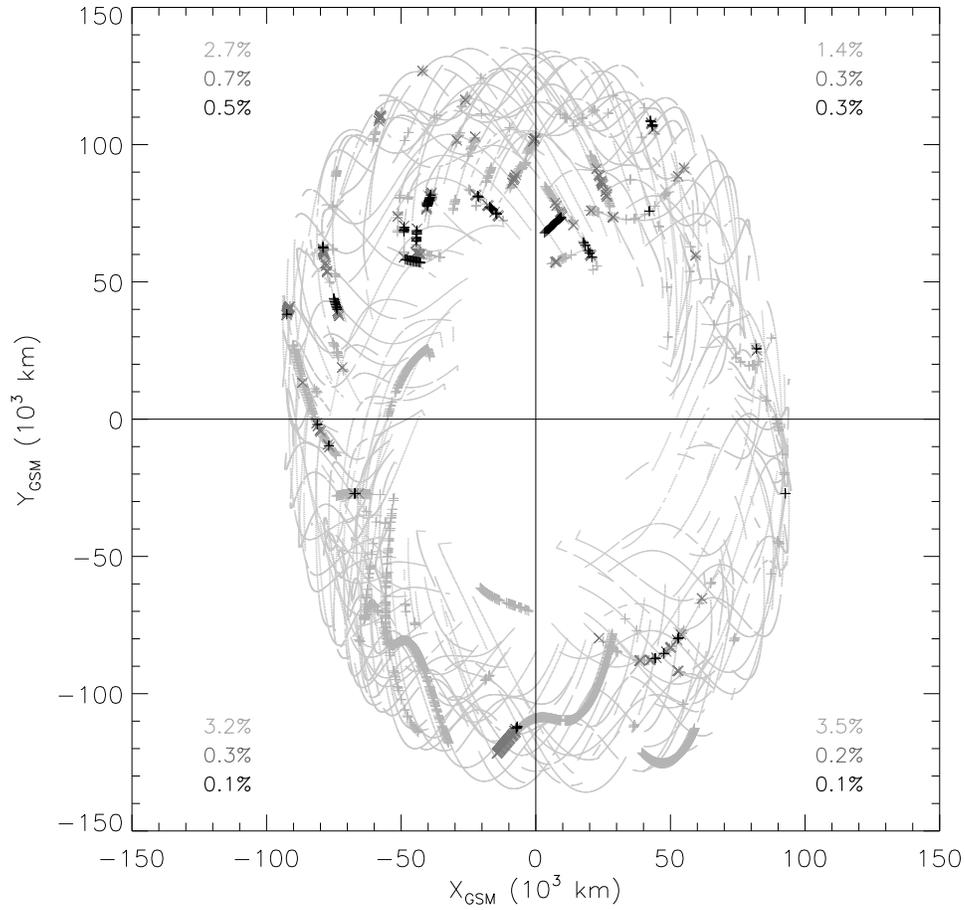}
\caption{Location in GSM coordinates when ACIS background is above the quiescent level.  Symbols and greyscale are as in Figure 1.  Projected in the $X_{GSM}-Y_{GSM}$ plane; the Earth is at the origin, the sun is in the direction $+X_{GSM}$, Earth's magnetic north pole is in the $+Z_{GSM}$ direction. }
\end{figure}

The frequency and strength of ACIS background flares seem to have a quasi-annual periodicity which could be linked with Chandra's position relative to the Earth's geomagnetic environment.  Chandra's orbit is nearly fixed in inertial space, so as the Earth orbits the sun, Chandra samples different regions of the magnetosphere which remains aligned with respect to the sun.  Figure~5 shows the locations in geocentric solar magnetic (GSM) coordinates when the ACIS background is 1-, 3- and 5-sigma above the quiescent level.  The numbers in the corners indicate the percentage of the time spent in each quadrant that the background was at a given flaring level.  The Chandra Radiation Model (CRM; see Blackwell et al. 2000) incorporates data from a number of sources to model the low-energy proton flux in Chandra's orbit and predicts higher proton fluxes in the midnight through dusk sector ($X_{GSM} < 0, Y_{GSM} > 0$) and minimum fluxes in the daytime through dawn sector ($X_{GSM} > 0, Y_{GSM} < 0$).  This asymmetry is certainly seen in the position dependence of the strongest ACIS background flares ($> 3\sigma$ over the quiescent level) but not for the more common weaker flares.  Since Chandra's orbit is inclined, a two-dimensional representation may not indicate fully which geomagnetic region Chandra is transiting.

\section{Conclusions}
The Chandra particle background is highly variable on both the long and short-term.  The quiescent background is due to to cosmic rays, while the flaring background may be associated with low energy protons in more local geomagnetic structures.  While some correlation is seen with altitude and geomagnetic position, more complete modeling of solar activity and the changing geomagnetic environment may be necessary for full understanding of background flares.

\acknowledgments We would like to thank our colleagues at MIT and the CXC especially Peter Ford.  ACE browse data was obtained from the ACE Science Center Web page (http://www.srl.caltech.edu/ACE/ASC/).  This work was supported by NASA contracts NAS8-37716, NAS8-38252 and NAS8-39073.

\end{document}